\pdfoutput=1
\documentclass{eas}
\usepackage{graphicx}
\usepackage[latin1]{inputenc}
\usepackage{amsmath}
\usepackage{color}
\usepackage{amssymb}
\begin{document}

\title{Gravitational Lensing Statistics by Galaxy Clusters with smoothness parameter depending on z} 
\author{Casta\~neda L.}\address{Observatorio Astron\'omico Nacional. Universidad Nacional de Colombia\\ leonardo@astro.uni-bonn.de}
\author{Valencia-D M.}\address{Grupo de F\'isica y Matem\'atica. Universidad Pedag\'ogica Nacional\\ mvalencia@uni.pedagogica.edu.co}
%
%
\begin{abstract}
Nowadays, Gravitational lensing is a fundamental tool to study the light propagation in the universe and the evolution of structures.  Here, diferent forms of the Santos \& Lima smoothness parameter as a funtion of cosmological redshift are used in order to describe various scenarios of evolution of structure. Its effects on the strong gravitational lensing SGL by  galaxy cluster dark halos are analized in LCDM, EdS and OCDM cosmologies.  The dark matter halos modeled by Navarro-Frenk-White singular profile and the Press-Schechter approximation is assumed to describe its distribution. We found that SGL probability by galaxy clusters becomes to be sensible to the evolution of structures scenario considered for sources farther than $z \approx 2$ and it strongly depends on the cosmology and the smoothness parameter actual value.
\end{abstract}
\maketitle
\section{Introduction}
\noindent The universe is locally inhomogeneous. This inhomogeneity can be expressed by the smoothness parameter $\tilde{\alpha}$ proposed by Dyer and Roeder in 1972 as a constant with values between 0 (all the matter clumped in structures) and 1 (strictly FRW case), which describes the uniformly distributed matter proportion.  In Strong Gravitational Lensing SGL this fact can be introduced via the eikonal approximation \cite{sef} finding that SGL probability depends on $\tilde{\alpha}$ \cite{csyr} \cite{leo}.  However, due to the structure formation process, $\tilde{\alpha}$ must vary on time and SGL must be affected. Here we show how is the variation of the SGL probability by dark matter halos DMH of galaxy clusters when different CDM scenarios of evolution of structure, represented by different forms of $\tilde{\alpha}(z)$, are considered. We assume spherically symmetric lenses to emphasize in the effect of the structure evolution, although recently has been shown that the ellipticity of the DMH affects strongly the SGL \cite{men}. The considered cosmologies are EdS ($\Omega_{M0}=1$,$\Omega_{\Lambda0}=0$), OCDM ($\Omega_{M0}=0.3$,$k=-1$) and LCDM($\Omega_{M0}=0.3$,$\Omega_{\Lambda0}=0.7$) with $h=0.72$ and $\Omega_B0=0.071$. In Sec. 2 the smoothness parameter and its introduction in the GL are described. In Sec. 3 the lens model and distribution mass function are described. In Sec. 4 the GL probability in the three cosmological models is calculated and in Sec. 5 the main conclusions are presented.

\section{Smoothness Parameter}
\noindent The smoothness parameter $\tilde{\alpha}$ used to describe the local matter distribution have to be inhomogeneous and anisotropic, but if an average value over concentric spheres centered in the observer is considered, it can be express only as a function of cosmological redshift.  This evolution of  $\tilde{\alpha}$  is related with the evolution of structure in the universe, so that, $\tilde{\alpha} \rightarrow 1$ for $z \gg 1$ to recover the homogeneity of the first stages of the structure formation processes, and while $z \rightarrow 0$, $\tilde{\alpha}$ approaches to the present value, which is unknown.  These requirements are full filled for the Santos \& Lima smoothness parameter \cite{syl}
\begin{equation}
	\tilde{\alpha}=\frac{\beta_0 (1+z)^{3\gamma}}{1+\beta_0(1+z)^{3\gamma}}
\end{equation}
\noindent where $\beta_0$ and $\gamma$ are dimensionless and can be used to obtain different scenarios of the evolution of structure. 
In eikonal approximation this parameter is related with the convergence of light bundles that are propagated in a region far from structures, this effect is know as focusing and it causes the change of the angular diameter distances ADDs. This means that the variation of $\beta_0$ and $\gamma$ gives different ADDs according to the Dyer- Roeder equation. In SGL distances between source-lens-observer are ADDs so, by this way, the effect of evolution of structure is introduce in the study of the probability to observe multiple imaged systems

\section{Gravitational Lensing by DMH of Galaxy Clusters}

\noindent The probability of SGL with multiple images depends on source and lens positions, distribution, the models used to describe them and the cosmological parameters.
The distribution function describes the comovil numeric density $n(z,M)$ of lenses, which in this case are DMH of galaxy clusters virialized at $z$ with mass in $(M,M+dM)$. The distribution function considered here is the Sheth-Tormen function \cite{syt}, which generalizes the Press-Schechter function in a wide range of CDM cosmologies.
\begin{equation}
				\frac {dn}{dM} = \sqrt{\frac{2aA^2}{\pi}} \, \frac{\rho_0}{M^2}\, \frac {\delta_c(z)}{\sigma (M)} \left\{ 1+ \left[ \frac{\sigma (M)}{\sqrt{a} \delta_c (z)}  \right] ^{2p}   \right\}  \left| \frac{d \ln \sigma}{d \ln M}  \right| \exp \left[ - \frac{a \delta^2_c(z)}{2 \sigma^2(M)} \right] 
\end{equation}
\noindent The best fit is when $a=0.707$, $p=0.3$ and $A= \approx 0.322$. Here $\rho_0 (\approx 2.78 \times 10^{11} \Omega_0 h^2 M_{\odot} \text{Mpc}^{-3})$ is the mean mass density at the present time, $\delta(z)$ represents the density contrast of a virialized object at $z$ extrapolated to lineal regimen $\delta_c(z)=\delta(t_{vir}) \left[ D_{+}(z=0)/D_{+} \right]$, with $D_{+}$ the lineal growing factor of desity fluctuations, which was calculated through Li \& Ostriker expressions \cite{lyo}. In (3.1) the mass variance of the fluctuations $\sigma(M)$ was obtained using Kitayama \& Suto approximation \cite{kys}.

\noindent The density profile of DMH of galaxy clusters is suitable described by NFW singular profile \cite{nfw}, which is independent on the primordial density fluctuations spectrum and on the cosmological parameters.  Inside the scalar radius $r_s$, the DMH density near the center goes as $\rho \sim r^{-1}$ and far form it as $\rho \sim r^{-3}$. Because of its singularity, this profile is used only to describe external regions.

\noindent The Lens Equation sets the mapping between the source plane and the lens plane, showing for each source position $y$, the position of its images $x$. In the case of NFW profile, this relation is
\begin{equation}
		y=x-C \, \,\frac{ g(x)}{x} \; \; \text{with }\; \; 
 g (x)= \ln \left( \frac{x}{2} \right)
\begin{cases}
\frac{2}{\sqrt{1-x^2}} \; \text{arctanh} \sqrt {\frac {1-x}{1+x}} & \; (\text{\textit{x} $<$ 1}) \\
1 & \; (\textit{x}=1)\\
\frac{2}{\sqrt{x^2-1}} \; \arctan \sqrt {\frac {x-1}{1+x}}  & \; (\text{\textit{x} $>$ 1})\\
\end{cases} 
\end{equation}
and $C \, \equiv \, 6 \,  \frac{H^2}{H_0 c} \, \frac{r_{OL} \, r_{LS}}{r_{OS}} \, \delta_s \, r_s$. The order of magnitude of  $C$ is $10^{-1}$ to 10, depending strongly on the critical sobredensity of the halo $\delta_s$.  The influence of cosmology and the structure formation scenario can be noticed by the presence of ADDs.  The relative maximum of (3.2) is the radius of caustic circumference $y_c$, which limits the area of the sky where a source must be placed in order to produce three images; this region is known as cross-section $\sigma$.  The cross-section depends on $\tilde{\alpha}$ through C.  As first approximation, it was considered $6 \,  \frac{H^2}{H_0 c} \approx 50$.

\section{Results}

\noindent The probability for a source at $z_S$ to undergo a strong lensing event with multiple images by the presence of a DMH of a galaxy cluster at $z_L$, is
\begin{equation}
		P(z_s)= \left\langle \mu\right\rangle_{z_S} \frac{c}{H_0} \int_0^{z_S}\frac{(1+z)\; \left\langle \mu\right\rangle_z^{-1} \; dz }{\sqrt{\Omega_{M0}(1+z)^3+\Omega_{\Lambda 0} + \Omega_{k0}(1+z)^2}} \int_{M_0}^{M_1}  dM  \frac{dn}{dM}\, {\sigma}r_s^2
\end{equation}
\noindent here $\left\langle \mu\right\rangle_z =\left[{r^2(z)}/{r_1^2(z)} \right]^{2}$, $r(z)$ is the ADD calculated from $z=0$ to $z=z$ using the smoothness parameter as a function of cosmological redshift and $r_1(z)$ is the ADD for a strictly FRW model. Evaluating the probability for the considered cosmological models (Fig. 1 a 3) it can be noticed that highest values are obtained in the LCDM case, and that the larger dispersion of the curves for $z \sim 2$ is got in EdS cosmology implying a greater effect of the evolution of structure in GL. In Fig. 3, the probabilities obtained assuming constant smoothness parameters $\tilde{\alpha}=1$ (upper green line) and $\tilde{\alpha}$ =0 (lower green line), are presented as reference.

\begin{figure}[tbh!] 
\begin{minipage}{0.45\textwidth}
\centerline{\includegraphics[bb=0 0 450
400,height=3.5cm,clip]{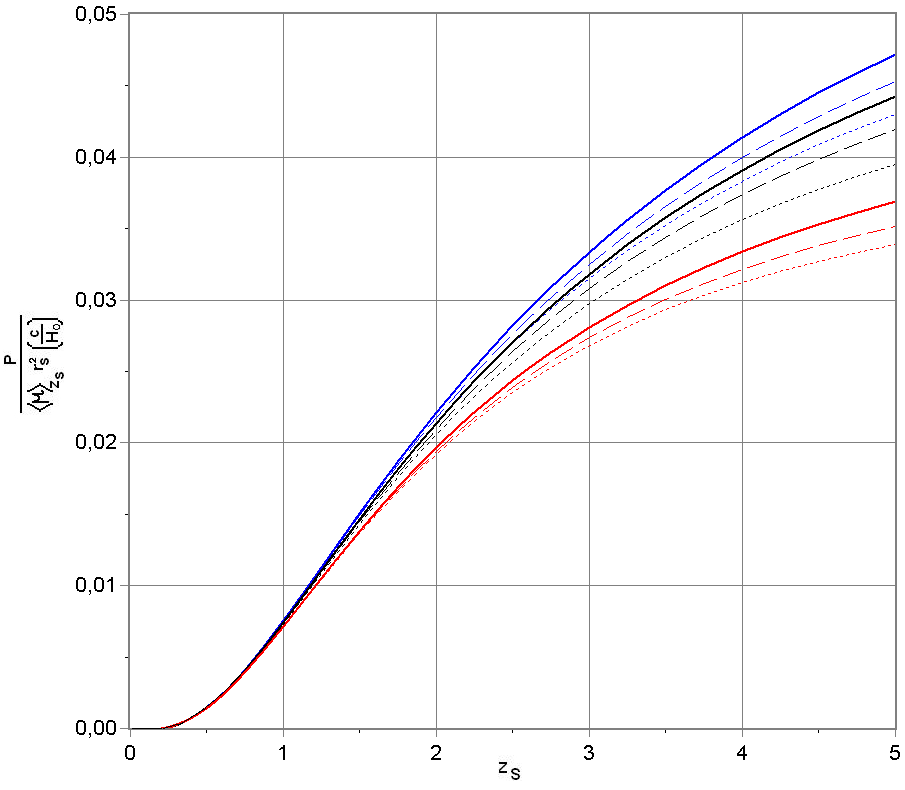}}
\caption{Probability in EdS.} 
\end{minipage} \; \;
\begin{minipage}{0.45\textwidth}
\centerline{\includegraphics[bb=0 0 450
400,height=3.5cm,clip]{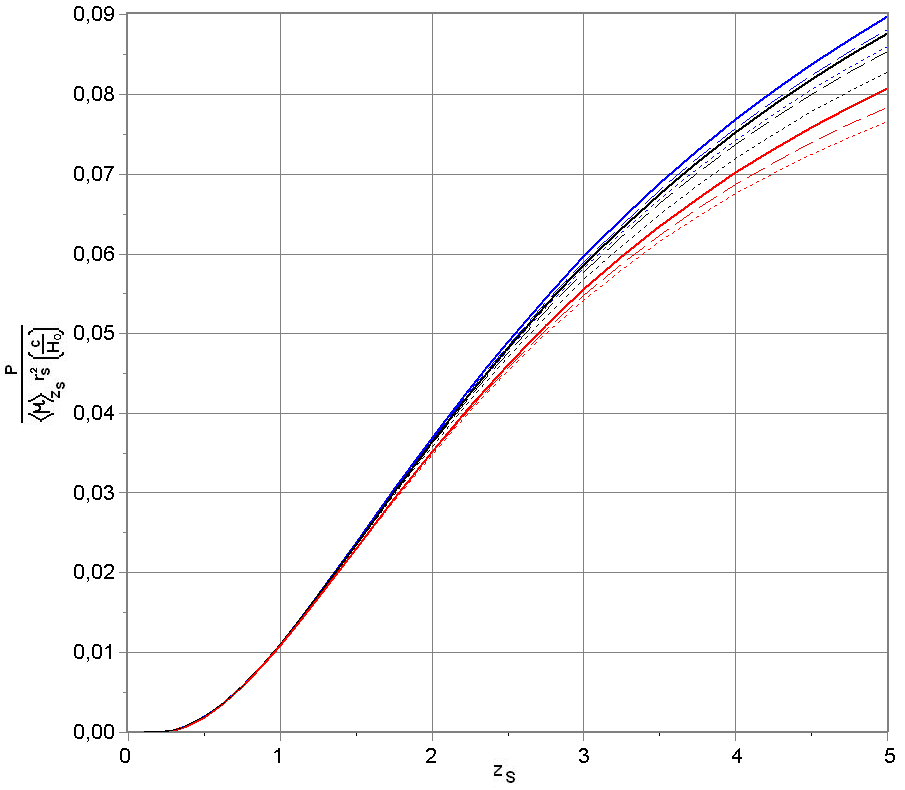}}
\caption{Probability in OCDM.}
\end{minipage}
\end{figure}

\begin{figure}[tbh] 
\begin{minipage}{0.45\textwidth}
\centerline{\includegraphics[bb=0 0 450
400,height=3.5cm,clip]{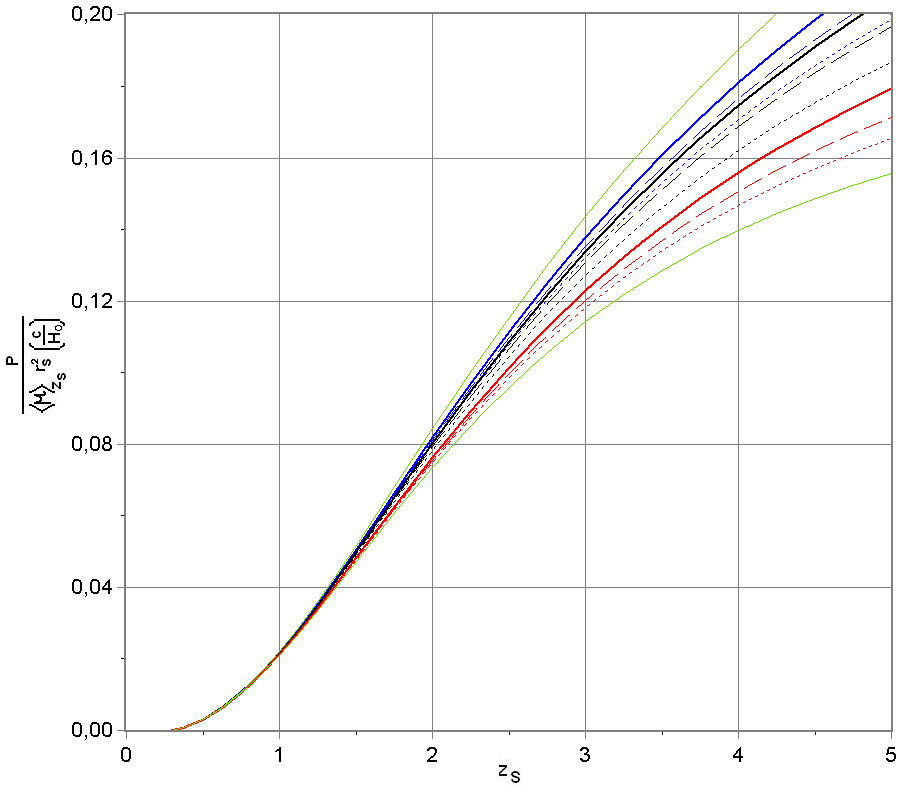}}
\caption{Probability in LCDM.} 
\end{minipage} 
\begin{minipage}{0.6\textwidth}
\centering
\begin{tabular}{|p{0.8cm}| c | c |} \hline
Trazos&$\beta_0$&$\gamma$\\ \hline		
\textcolor{blue}{\rule{0.6cm}{1pt}}&1.0&0.7\\ \hline
\textcolor{blue}{\rule{0.2cm}{1pt} \rule{0.2cm}{1pt}}&1.0&0.5\\ \hline
\textcolor{blue}{\rule{1.2pt}{1.2pt} \rule{1.2pt}{1.2pt} \rule{1.2pt}{1.2pt} \rule{1.2pt}{1.2pt}}&1.0&0.3\\ \hline
{\rule{0.6cm}{1pt}}&0.5&0.7\\ \hline
{\rule{0.2cm}{1pt} \rule{0.2cm}{1pt}}&0.5&0.5\\ \hline
{\rule{1.2pt}{1.2pt} \rule{1.2pt}{1.2pt} \rule{1.2pt}{1.2pt} \rule{1.2pt}{1.2pt}}&0.5&0.3\\ \hline
\textcolor{red}{\rule{0.6cm}{1pt}}&0.1&0.7\\ \hline
\textcolor{red}{\rule{0.2cm}{1pt} \rule{0.2cm}{1pt}}&0.1&0.5\\ \hline
\textcolor{red}{\rule{1.2pt}{1.2pt} \rule{1.2pt}{1.2pt} \rule{1.2pt}{1.2pt} \rule{1.2pt}{1.2pt}}&0.1&0.3\\ \hline
\end{tabular}
\end{minipage} 
\end{figure} 

\section{Conclusions}

\noindent The probability to observe a SGL with production of multiple images by a DMH of a galaxy cluster depends strongly on the cosmological model considered and the parameters used to describe the lens e.g. $\delta_c$.  In other hand, for sources at $z>2$ the difference in the probability increases, which is higher when the considered current value of smoothness parameter $\tilde{\alpha}(z=0)$ is higher and when scenarios with later formation of systems are assumed (high values of  $\gamma$ ). 


\begin{thebibliography}{99}
\bibitem[2000]{leo} Casta\~neda, L. 2000, Master Thesis. Universidad Nacional de Colombia
\bibitem[2005]{csyr} Covone, G., Sereno, M. and de Ritis, R. 2005, MNRS, 375, 773
\bibitem[1996]{kys} Kitayama, T. and Suto, Y. 1996, ApJ, 469, 480
\bibitem[2002]{lyo} Li, X. and Ostriker, J. 2002, ApJ, 566, 652
\bibitem[2007]{men} Meneghetti \etal\ 2007, A\&A, 461, 25
\bibitem[1997]{nfw} Navarro, J.F., Frenk, C.S. and White, S.D.M. 1997, ApJ, 490, 493
\bibitem[2006]{syl} Santos, R.C. and Lima, J.A. ArXiv astro-ph/0609129
\bibitem[1992]{sef} Schneider P., Ehlers, J. and Falco, E.E. 1992, {\it ``Gravitational Lenses''}, Springer-Verlag 
\bibitem[1999]{syt} Sheth, R.K. and Tormen,G. 1999, MNRS, 308, 119
\end{thebibliography}
\end{document}